\documentclass[11pt,a4paper]{article}
\usepackage{graphicx}
\usepackage{natbib}
\usepackage{hyperref}

\begin{document}

\title{Enabling the Sustainable Space Era by Developing the Infrastructure for a Space Economy} 

\author{\textbf{Guillem Anglada-Escud\'e$^{1,2}$}
\\ 
\small    $ $\\
\small$^{1}$Ramón y Cajal research fellow,
\small         Institut de Ciències de l’Espai - CSIC\\
\small         08193, Barcelona, Spain, 
\small    $ $\\
\small    $ $\\
\small   $^{2}$Institut d'Estudis Espacials de Catalunya (IEEC),\\ 
\small         08034, Barcelona, Spain             
}



\date{Accepted for publication in EXPA: Aug 20, 2021}

\maketitle

\begin{abstract}
The world is changing fast, and so is the space sector. Planning for large scientific experiments two decades ahead may no longer be the most sensible approach. I develop the argument that large science experiments are becoming comparable to terrestrial civil infrastructures in terms of cost. As a result, these should incorporate plans for a return on investment (or impact, not necessarily economic), require a different approach for inter-division coordination within the European Space Agency (ESA), and a broader participation of all society stakeholders (civil society representatives, and the broader public).

Defining which experiments will be relevant two decades ahead adds rigidity and quenches creativity to the development of cutting edge science and technology. This is likely to discourage both senior and earlier career professionals into supporting such long-term (and often precarious) plans. A more sensible strategy would be increasing the rate of smaller well understood experiments, engage more society sectors in the development of a truly space-bound infrastructure, and formulate a strategy more in tune with the challenges faced by our society and planet. We argue that such strategy would lead to equally large -even larger- scale experiments in the same time-scale, while providing economic returns and a common sense of purpose. A basic but aggressive road map is outlined.

\textbf{Keywords: } \textit{Space infrastructure, Economy, Public engagement, Science Communication}
\end{abstract}

\section{Introduction and vision}
\label{intro}
The true space era can only begin once we learn how to live and thrive in space without further straining Earth. That is, once we are able to start a space-based economy. Moreover, our planet is one of many places in space, so by learning how to start self-sustained human activities in space, we will advance towards a zero footprint economy back on Earth, which is a worthy legacy for the future generations. That is, space provides the chance to improve efficiency over ground based solutions (e.g. satellite telecommunications as opposed to wired networks), and move certain highly contaminant and planet surface consuming activities (e.g. energy production, semiconductor manufacture) off-world. This can be done in a global community context by engaging and coordinating experts outside the traditional space academic fields; training new professionals, organizing workshops and conferences, and developing discussion forums and citizen science initiatives.

The synergies between this vision and current grand challenges on Earth are numerous. For practical economic sustainability reasons, basic and technology research on new processes in space cannot be left to the private sector only. By publicly supporting new methodologies without the need to satisfy the immediate economic incentive, we can develop breakthroughs on a broad range of disciplines (e.g., spin-offs of space race, computing sciences, manufacture, material sciences). A space-based economy implies new costs \& value systems detached from Earth’s one. That is, terrestrial launches may be devoted to deliver advanced high-technology components, while basic materials can be supplied directly from orbit. For example, due to launch costs one kilogram of fuel in orbit currently has almost the same cost as one kilogram of electronic components. The production of \textit{simple materials} from space (e.g. fuels, mechanical components) seems feasible within the next few years. The availability of these would dramatically reduce cost in Earth currency and effort, enabling more missions and/or larger scale ones.

Space endeavours are traditionally associated with governments, elites and scientists. As a result, the citizens are usually relegated to bystander roles, and a lot of creative power remains latent. My vision is that we need to break this stand-off and tap into the collective intelligence of societies and its citizens. In addition to partnerships with traditional actors, the new space exploration era beyond the next decade must enable citizen participation (companies, individuals, schools, associations, towns) through the use of cooperation, co-creation, and citizen science tools. Last but not least, the inspirational value of space coupled to  the new systems of cost and value discussed earlier will be a powerful playground to attract young minds into pure, applied and even social sciences.

Beyond the pragmatic use of developing a space-based economy; doing this will be essential to reach significant advances in all traditional basic scientific endeavors. That is, once we are unbound to Earth limitations, we can proceed to truly revolutionary advances such as
\begin{itemize}
\item Gigantic $>$20m class telescopes, possibly reaching 100~m apertures and beyond (using interferometers), enabling extremely high resolution imaging beyond imaginable today. This includes direct imaging and precision spectroscopy of nearby exoplanets, many of them potentially habitable, to the point where details on their surface (continents, ice-caps, oceans, cloud bands) can begin to be resolved \citep{Labeyrie2016}.
\item Arrays of antennas for radio observations on the far side of the moon\citep{Bandyopadhyay2018} and in deep space will enable access to a wide range of astrophysical phenomena, the very early universe (dark ages between cosmic microwave background, and first stars), and extreme physics (black hole high resolution imaging around numerous nearby galaxies, and even galactic stellar black holes and neutron star objects).
\item Direct easy access to geologic records in other solar system bodies (moon, Mars, asteroids and beyond) will enable access to pristine records of the Solar System\citep{Matthewman2015} at a completely different level of measurement and human experience.
\item Astrobiology experiments in space \citep{Spahr2020,Yamagishi2014}) and other relevant Solar System environments, including the Moon, Mars, asteroids, interplanetary space and beyond. Current experiments on how life would develop in space, and how complex chemistry occurs are broadly limited to a handful of expensive experiments in the ISS and ground-based laboratory simulators
\end{itemize}
\noindent The scale of truly revolutionary next generation large scientific facilities is at the level of large civil infrastructures, such as the Panama channel upgrade \citep[$\sim$15 Bn~EUR][]{panama2015}, or the high speed train HS2 in the UK \citep[$>$50 Bn EUR][]{Oakervee2020}. At this level of investment, there has to be -at least- some prospect of immediate societal and economic return. Moreover, since large monolithic space experiments (launch and forget approach) are very vulnerable single-point failures, costs are likely to ramp up compared to current estimates as a result of risk mitigation matters, resulting in diminishing scientific returns per EUR. Therefore, when looking at what would be relevant three decades from now, it is essential equating the (likely) evolution of world economics and its geopolitics.

In this article I advocate for an aggressive development of the technologies and infrastructures needed to develop a space-based economic and industrial base, and plan the large science experiments in the pre-2050 time-frame accordingly. 

\section{Possible plan outline}
\subsection{Short-term (2020-2025)}
Start a vigorous multidisciplinary program to attract professionals from traditionally non-space disciplines. These include (but it is not limited to) architects, chemists, industrial engineers, manufacturing engineers, material scientists, mining engineers, artificial intelligence experts, sociologists, biologists, economists,\ldots just to mention a few. Professional societies, creation of professional networks (e.g. EU’s European Innovative Training Networks\footnote{\url{https://ec.europa.eu/research/mariecurieactions/actions/research-networks_en}, accessed Nov.~15 2020}, EU COST actions\footnote{\url{https://www.cost.eu/}, accessed Nov.~15 2020}) shall be encouraged and supported by ESA as the different communities are unlikely to engage in professional interactions without encouragement.

\subsection{Mid-term (2025-2035)}
The task of starting a space-based economy is daunting, and a dynamic road map --that allows for regular revisions-- would need to be organised first. During this time interval a program of multidisciplinary conferences and workshops endorsed or nurtured by ESA should enable first contacts with the different communities so synergies can appear. Similarly, these need to be coupled to both governments (member states, but also EU), and the broader society. Any space based endeavour for either scientific or economic activities will require massive amounts of public support. A professionally and strategically developed communication/dissemination plan with the different stakeholders should be seriously considered as an integral part of it, and cannot be underestimated. In this sense, the conversation with stakeholders must be bidirectional, and enabling the broader society to be involved in decision making so the right balance is achieved between economic and scientific advancement.

\subsection{Long-term (2035-2050)}
As a specific milestone, a research station focused on technological capabilities development rather than basic science should be deployed on the surface of the Moon, and eventually develop it into a first industrial facility beyond Earth. As a collateral consequence of a crewed Lunar base, scientific research can immediately take advantage of a crewed base for sample collection and analyses, geologic studies, construction of a Moon based observatory (small telescopes and antenna arrays first), and astrobiology \& human physiology studies. 

The main goal of this first facility would be testing and developing all the In-Situ-Resource-Utilization (ISRU) and manufacturing processes to enable self sustainability, and to service Earth orbit infrastructure. After the first deployment years, the \textit{base} can be operated as a service provider for experiments (public, governmental or private) approaching the model of Antarctica research stations and/or large observatories on Earth \footnote{See for example Keck Observatory\citep{keck} as an example of a public/private partnership scientific organization in \url{https://www.keckobservatory.org/about/keck-observatory/}, and documents therein. Accessed Nov~15 2020.}. Some agencies, most remarkably NASA\citep{artemis}\footnote{\url{https://www.nasa.gov/specials/artemis/}, accessed Nov~15, 2020} and the Chinese National Space Agency\citep{chinaMoon}\footnote{\url{http://www.clep.org.cn/}, accessed Nov~15, 2020}, are fully engaged with establishing a Lunar-based facility towards the end of the 2020's. ESA also has layout plans to develop Lunar exploration capabilities \citep[European Large Logistics Lander, ][]{esaLander}, and it has an initial plan for development of ISRU\citep{esaISRU}, but it is remarkably less defined, (and funded) than those of competing agencies. In all cases, there is not a clear connection of this capability development with the long-term large science experiments as those discussed in the ESA Voyage 2050 exercise. That is, most large space missions (aka.~ESA L-class missions) still assume a framework based on Earth launches only, with only a few shy attempts to propose novel concepts using the Moon as a platform for astronomical observations \citep[see][as an exceptional example]{owlMoon}. 

\subsection{Beyond 2050}
Beyond 2050, one could imagine that a self-sustaining self-replicating infrastructure is also deployed on Mars\citep{musk}. Again, a wide range of experiments can begin immediately, but more importantly, having a destination for human activities enables the beginning of economics between Earth orbit and Mars; further boosting space and Lunar activities, and the beginning of the the systematic exploration/exploitation of asteroids. Achieving self-replicating capabilities ensures that expansion in Mars and into the Solar system becomes resilient against changing circumstances on Earth economics. At that point, and given that there will be surpluses, the space infrastructure should be able to start feeding resources back to Earth (especially Earth orbit servicing); while expanding human activities in the Solar System. Next generation space mega-projects ($>$100~m telescope, $>10^6$~m baseline radio-arrays, etc.) can also start at this point at a fraction of current cost estimates in Earth currency.

\section{Conclusions and discussion}
Many urgent challenges facing us, and the uncertainties on the mid term stability of the world economy \footnote{Recurrent crisis and financial have a direct impact on support for basic science, see ESA/JAXA SPICA case as a recent example (abandoned after many years of community development), https://sci.esa.int/web/cosmic-vision/-/spica-no-longer-candidate-for-esa-s-m5-mission-selection}, I find it might be very unwise to maintain the current approach of planning for one shot L--class missions which -at best- will satisfy a small part of the already small scientific communities. Moreover, these large missions are bound to suffer from uncertainties of the world’s geopolitics too. Large-scale projects  represent non-negligible (multi)state level efforts and, in terms of funding, they will unavoidably compete for funds with crucial civil infrastructures when discussed at government level\footnote{see for example, the rampaging costs and delays of the James Webb Space Telescope (from one Billion USD to current ten Billion USD price tag), the several near cancellations, and the resulting political tensions generated within the US, \url{https://www.theverge.com/2018/8/1/17627560/james-webb-space-telescope-cost-estimate-nasa-northrop-grumman}. Source : The Verge website, Accessed Nov~15, 2020}. A much better strategy would be proposing large investments to develop capabilities with a much more clear return potential, at least in principle (e.g. proposing a Lunar base for support to Earth orbit infrastructure). On the other hand, incremental small scale experiments with shorter development time-scales (such as M and S--class missions, say $<1$ billion EUR scale) would be less sensitive, and able to proceed at an increased rate thanks to the reduction of launch costs already happening.

There are world-wide efforts devoted on how to solve the climate crisis/catastrophe, especially among the younger generations\footnote{The Greta Thunberg and Fridays for Future phenomenon would be the most notorious examples, \url{https://fridaysforfuture.org/}}. After interacting with stakeholders beyond traditional academic backgrounds\footnote{see \url{https://www.debatingeurope.eu/2020/07/07/should-europe-set-up-a-moon-base/}, Accessed Nov.~15, 2020}, I find that long term support for large scale space scientific activities is unlikely, and possibly not morally justifiable. In this sense, when planning for large-scale scientific endeavours beyond the 2030 horizon, these need to be part of the \textit{solutions} to the world’s problems.

In terms of missions to implement this vision, I envisage a road map starting with community gathering activities of multidisciplinary nature, supported by seed level research grants and programs. We can then establish a first Moon infrastructure (~2030’s) dedicated mostly to develop industrial \& manufacturing capabilities in space with the sight set on Mars. By $\sim$2040, we may be able to deploy a self-sustainable and self-replicating infrastructure on Mars. This will open up the space between Earth and Mars to human economic activities, at which point the nascent space economy can start feeding back to Earth’s; and a new generation of space mega-projects (current concepts, or yet to be conceived) can begin.

\section{Addendum} 
\textit{Additional considerations submitted after Voyage 2050 workshop, October 2019, Madrid, Spain}. Having exchanged opinions and ideas with colleagues and other representatives of science related ESA activities in space (Directorate of Earth observations, and Human and Robotic Exploration programs) and other stakeholders (forums with participants at EU level institutions), I have a few additional practical comments to make

\begin{itemize}
    \item 
It seems that there is a lack of horizontal coordination between ESA directorates. While the existence of separate divisions is understandable from an operational point of view, a more coordinated long term strategy would seem advisable. I consider this especially crucial when drafting long term strategic of the agency beyond the next decade. In this sense, exercises similar to \textit{ESA Voyage 2050} with contributions from inter-division participants and communities would seem a sensible course of action\footnote{The European Logistics Lunar Lander (EL3) program would be a first step in this strategy, but as discussed earlier; it has a much more modest scope than the USA or China counter parts. The EL3 program was announced after the initial submission of this paper.}.

\item 
Along the same lines, I strongly emphasize the recommendation that ESA shall work hard to improve its interaction with other societal stakeholders, especially towards sectors (academic and industrial) not traditionally involved in space related activities. This crucial (and often underappreciated) to keep long term support and funding to space science activities. Basic science must proceed, but the conversation must exist between sectors to ensure it remains meaningful to the citizens.

\item 
Discussing about large experiments (i.e. L--class missions) two decades ahead seems unnecessary and, in the light of past experience of cancellations (Eddington/ESA, cancelled 2003), rescopings (BepiColombo) and massive delays (LISA, initially planned in the 2020-2025 interval, currently scheduled at 2034), it is likely to be discouraging to a wide range of professionals and scientists (both senior and early career researchers). Instead of this, it is clear that the scientific community would be keen on having more frequent (possibly smaller) experiments, which in turn would enable a quicker technology development, eventually enabling for breakthroughs as well. The interest for more frequent medium class experiments can be seen in the submission statistics for the M-class mission calls in the ESA Cosmic Vision context, where more than twenty mission concepts have been consistently proposed at each round\footnote{See call for M5 mission slot, \url{https://www.esa.int/Science_Exploration/Space_Science/ESA_selects_three_new_mission_concepts_for_study}}. In this sense, a recommendation would be running missions at a faster rate (lowering cost, accepting higher risk and acquiring more experience); and transform the science program into a more adaptable entity with shorter development times.

\item More particular to the white paper topic, but closely related to the aforementioned points, it would be exciting to see the ESA Human \& Robotic Exploration and the ESA Science directorates preparing more ambitious and coordinated actions together. Along these lines, the possibility of developing activities and observatories on the Moon, should be more seriously considered as a natural point of contact for both in the next two decades and beyond.
\end{itemize}

\small \noindent \textbf{Acknowledgements.} The author is supported by the Ram\'on Y Cajal fellowship RYC-2017-22489 from Ministerio de Ci\'encia, Innovaci\'on i Universidades from Spain. The author thanks the organizers \textit{ESA Voyage 2050} exercise for providing the opportunity to the community to expressing their views in an open environment. \textit{Note added in proof.} This \textit{article} and \textit{addendum} were submitted before the COVID19 crisis, further emphasizing the need of the development of a science \& technology road map that is both robust and flexible against the developments of the world.
%
\section*{Conflict of interest}
The authors declare that they have no conflict of interest.

\bibliographystyle{spbasic}      
\bibliography{biblio}   

\end{document}